\documentstyle[12pt]{article}

\begin{document}

\title{Cosmic Equation of State and Advanced LIGO Type Gravity Wave Experiments}
\vspace{0.6 cm}
\author{ {\sc Marek Biesiada} \\
{\sl Department of Astrophysics and
 Cosmology, } \\
 {\sl University of Silesia}\\
 {\sl Uniwersytecka 4,  40-007 Katowice, Poland}  \\
 mb@imp.sosnowiec.pl}

 \date{}

\maketitle \vfill

\begin{abstract}
\noindent
Future generation of interferometric gravitational wave detectors is hoped to
provide accurate measurements of the final stages of binary inspirals.
The sources probed by such experiments are of extragalactic origin and the
observed chirp mass is the intrinsic chirp mass multiplied by $(1+z)$ where $z$
is the redshift of the source. Moreover the luminosity distance is a direct
observable is such experiments. This creates the possibility to establish a
new kind of cosmological tests, supplementary to more standard ones.

Recent observations of distant type Ia supernovae light-curves suggest that the
expansion of the universe has recently begun to accelerate. A popular
explanation of present accelerating expansion of the universe is to assume that
some part $\Omega_Q$ of the matter-energy density is in the form of dark
component called ``the quintessence'' with the equation of state $p_Q =
w \rho_Q$ with $w \geq -1.$
In this paper we consider the predictions concerning observations of binary
inspirals in future LIGO type interferometric experiments assuming a
``quintessence cosmology''. In particular we compute the expected redshift
distributions of observed events in the a priori admissible range of parameters describing the
equation of state for the quintessence. We find that this distribution has a
robust dependence on the cosmic equation of state.

\end{abstract}

\section{Introduction}

Recent distance measurements from high-redshift type Ia supernovae
\cite{Riess98,Perlmutter99a} suggest that the universe is presently accelerating its expansion.
A popular explanation of this phenomenon is to assume that
considerable amount $\Omega_Q \approx 70 \%$ of the matter-energy density is in the form of dark
component called ``the quintessence'' characterised by the equation of state $p_Q = w \rho_Q$ with
$w \geq -1$ \cite{Caldwell98,Chiba98,Turner97}.
The evidence for spatially flat universe, reinforced by recent cosmic microwave background
experiments BOOMERANG and MAXIMA \cite{deBernardis,Hanany} calls for an extra unclustered
dark component. Within the standard cold dark matter (CDM) scenario only about
$0.2 < \Omega_{CDM} <0.4$ can be clustered in order to be in agreement with galactic rotation
curves, abundance of galaxy clusters, gravitational lensing or large scale velocity fields.
Moreover the accelerated expansion of the universe can be achieved with extreme forms of matter.
Hence this extra component (quintessence) should be similar to the cosmological constant but
is allowed to have its own temporal dynamics.
Many current  models of dark matter in general and of quintessence in particular \cite{Frieman},
invoke the concepts from particle physics.
Particle physics, however, gives little guidance as to concrete models of quintessence.
Therefore it has been proposed in \cite{Nakamura99} that future supernova surveys may
allow reconstructing the quintessential equation of state.
In this paper we shall contemplate the feasibility of constraining the cosmic equation of
state from the gravitational wave experiments in a similar vein as proposed in \cite{Nakamura99}.

Laser interferometric gravitational wave detectors developed under the projects
LIGO, VIRGO and GEO600 are expected to perform a successful direct detection of
the gravitational waves. Inspiralling neutron star (NS-NS) binaries are among
the most promising astrophysical sources for this class of experiments
\cite{Thorne}.
Besides quite obvious benefits from seeing gravitational waves ``in flesh'' and
providing valuable information about dynamical processes leading to their
generation inspiralling binaries have one remarkable feature.
Namely, the luminosity
distance to a merging binary is a direct observable quantity easy to obtain
from the waveforms. This circumstance made it possible to contemplate
a possibility of accurate
measurements of cosmological parameters such like the Hubble constant, or
deceleration parameter \cite{ChernoffFinn,Markovic,Krolak}.
In particular it was pointed out by Chernoff and Finn \cite{ChernoffFinn}
how the catalogues of inspiral events can
be utilised to make statistical inferences about the Universe.
In the similar spirit we discuss in this paper the possibility to constrain
the quintessence equation of state from the statistics of inspiral gravitational wave events.

\section{Cosmological model}

We shall consider a class of flat quintessential cosmological
models. The spatially flat Universe has recently received a considerable
observational support \cite{Max} from the measurements of the position of first acoustic
peak at $l \approx 200$ in baloon experiments BOOMERANG \cite{deBernardis} and MAXIMA \cite{Hanany}.
This class is parametrized by two quantities: $\Omega_0$ and
${\Omega}_Q,$ where $\Omega_0 = \rho / \rho_{cr} =
\displaystyle{\frac{8 \pi G \rho_0}{3 H^2_0}}$ denotes the current matter
density as a
fraction of critical density for closing the Universe, $\Omega_Q$
is analogous fraction of critical density contained in the quintessence and
these two sum up to the value one.
The equation of state for the quintessence is assumed in a standard form:
$p = w \rho$ where $w \ge -1.$ This form of the equation of state is very
general in the sense that it contains the well known constituents of the
universe as special subclasses. For example $w=-1$ corresponds to the
cosmological constant $\Lambda$, $w=0$ -- the dust matter, $w=-1/3$ -- cosmic strings and $w=-2/3$ the domain walls.

Non Euclidean character of the space-time is reflected in distance measures.
For the introduction to observational cosmology and the problems of distances
in non-euclidean spaces (see e.g. \cite{Peebles}).
In order to fix the notation for further use,
let us introduce an auxiliary quantity ${\cal D}(z)$:
\begin{equation} \label{D}
{\cal D}(z) = \sqrt{\Omega_0(1+z)^3 + \Omega_Q (1+z)^{3(1+w)} }
\end{equation}
As it is well known, one can distinguish three types of distances:

(i) proper distance:
\begin{equation} \label{proper}
d_M(z) = \frac{c}{H_0}\;\int_0^z \frac{dw}{{\cal D}(w)} =
\frac{d_H}{h}\;\int_0^z \frac{dw}{{\cal D}(w)} =: \frac{d_H}{h} {\bar d}_M(z)
\end{equation}

(ii) angular distance:
\begin{equation} \label{angular}
d_A(z) = \frac{1}{1+z} d_M(z) = \frac{1}{1+z} \frac{d_H}{h} {\bar d}_M(z)
\end{equation}

(iii) luminosity distance:
\begin{equation} \label{luminosity}
d_L(z) = (1+z) d_M(z) = (1+z) \frac{d_H}{h} {\bar d}_M(z)
\end{equation}

As usually $z$ denotes the redshift, $h$ denotes the dimensionless Hubble
constant i.e. $H_0 = h \times 100 \; km/s\,Mpc$ and $d_H = 3.\times
h^{-1}\;Gpc$ is the Hubble distance (radius of the Hubble horizon).
The quantities with an overbar have been defined by factoring out the
dependence on the Hubble constant from respective quantities.
In the further discussion we will explore the following models:
$$
(\Omega_0,\;\Omega_Q) = \{ (0.2,0.8); \;
(0.3,0.7); \; (0.4,0.6) \}
$$
with the $w$ coefficient equal to $w = \{0,-0.2,-0.4,-0.6,-0.8,-1. \}$

>From the observational point of view in the light of constraints from
large scale structure and cosmic microwave background anisotropies, the 95\%
confidence interval estimates give $0.6 \le \Omega_Q \le 0.7$ and $-1. \le w <
-0.6$ \cite{Efstathiou99,Perlmutter99b}.

However we retain the full spectrum of a priori possible quintessential  equations of
state in order to illustrate the discriminating power of the gravitational
wave data discussed in this paper.

\section{Redshift distribution of observed events}

The gravity wave detector would register only those inspiral events for which
the signal-to-noise ratio exceeded certain threshold value $\rho_0$
\cite{ChernoffFinn,Finn} which
is estimated as $\rho_0 = 8.$ for LIGO-type detectors.
An intrinsic chirp mass
${\cal M}_0= \mu^{3/5} M^{2/5}$,where $\mu$ and $M$ denote the
reduced and total mass, is the crucial observable quantity describing the
inspiralling binary system.
The observed chirp mass ${\cal M}(z) = (1+z) {\cal M}_0$ scales with the
redshift and therefore can be used to determine the redshift to the source
(there is strong evidence that the mass distribution of neutron stars in binary
systems is sharply peaked around the value $1.4\;M_{\odot}$). Because the
luminosity distance of a merging binary is a direct observable easily read off
from the waveforms one has a possibility to determine the
precise distance -- redshift relation and hence to estimate the Hubble constant
\cite{Finn,Biesiada}.
For a given detector and a source the signal-to-noise ratio reads \cite{Finn}:
\begin{equation} \label{rho}
 \rho(z) = 8 \Theta \frac{r_0}{d_L(z)} \left(\frac{{\cal M}(z)}{1.2\,M_{\odot}}
\right)^{5/6} \zeta(f_{max}),
\end{equation}
where $r_0$ is a charateristic distance scale, depending on detector's
sensitivity,
$r_0 \approx 355\;Mpc$ for advanced LIGO detectors, $d_L$ is the luminosity
distance to the source,
$\zeta(f_{max})$ is a dimensionless function describing the overlap of the
signal with detector's bandwidth.

The adiabatic inspiral signal terminates when the
binary system reaches the innermost circular orbit (ICO). The corresponding
orbital frequency is $f_{ICO}$ and $f_{max}$ corresponds to observed
(i.e. redshifted) $f_{ICO}$
\begin{equation} \label{f_max}
f_{max} = \frac{f_{ICO}}{1+z} = \frac{710\:Hz}{1+z}
\left(\frac{2.8\;M_{\odot}}{M} \right),
\end{equation}
so $f_{max} \sim 710 \;Hz$ for neutron star binaries. It is argued that
$\zeta(f_{max})\approx 1$ for
LIGO/VIRGO interferometers \cite{ChernoffFinn,Finn}.\\

Let us denote  by ${\dot n}_0$ the local binary
coalescing rate per unit comoving volume. One can use "the  best guess" for
local rate
density  ${\dot n}_0 \approx 9.9 \; h \;10^{-8} \;\;Mpc^{-3}yr^{-1}$
as inferred from the three observed binary pulsar systems that will coalesce in
less than a Hubble time \cite{Phinney}. \\
Source evolution over sample is usually parametrized by
multiplying the coalescence rate by a factor $\eta(z) = (1+z)^D$,
i.e. ${\dot n} = {\dot n}_0 \; (1+z)^2 \; \eta(z)$ where the
$(1+z)^2$ factor accounts for the shrinking of volume with $z$
and the time dilation of burst rate per unit time. The
cosmological origin of gamma-ray bursts (GRBs) has been confirmed
since discoveries of optical counterpart of GRB 970228
\cite{Groot} and the measured emission-line redshift of $z=0.853$
in GRB 970508 \cite{Metzger}. It has also been known for quite a
long time that cosmological time dilation effects in BATSE
catalogue suggest that the dimmest sources should be located at
$z \approx 2$ \cite{Norris}. Consequently several authors tackled
the question of source evolution in the context of gamma-ray
bursts. Early estimates of \cite{Dremer} and Piran \cite{Piran}
indicated that BATSE data could accommodate quite a large range
of source density evoultion (from moderate negative to positive
one). Later considerations by Horack et al. \cite{Horack}
indicated that if $z=2$ is indeed the limiting redshift then a
source population with a comoving rate density $n(z) \sim
(1+z)^{\beta}$ with $1.5 \le \beta \le 2$ is compatible with
BATSE data. Later on Totani \cite{Totani} considered the source
evolution effects and based his calculations on the realistic
models of the cosmic star formation history in the context of
NS-NS binary mergers.  Comparison of the results with BATSE
brightness distribution revealed that the NS-NS merger scenario
of GRBs naturally leads to the rate evolution with $2 \le \beta
\le 2.5$. We shall therefore take the source evolution effetcs
into account in our further considerations. One should stress, however that 
NS-NS merger scenario is by no means the unique explanation of gamma-ray bursts. 
Recently the so called collapsar model became popular \cite{collapsar}. The idea that at least some of gamma-ray 
bursts are related to the deaths of massive stars is supported by the observations 
of afterglows in GRB 970228 and GRB 980326 \cite{afterglows}. Therefore we will not 
prefer any specific value of evolution exponent $D$ but instead we will try to 
illustrate how strongly and in which direction does the source evolution affect our 
ability to discriminate between different quintessential equations of state.

The relative
orientation of the binary with respect to the detector is described by the
factor $\Theta$. This complex quantity cannot be measured nor
assumed a priori. However, its probability density averaged over binaries and
orientations has been calculated \cite{Finn} and is given by a simple
formula:
\begin{eqnarray} \label{P_theta}
P_{\Theta}(\Theta) &=& 5 \Theta (4 - \Theta)^3 /256, \qquad {\rm if}\;\;\;
0< \Theta < 4  \\
P_{\Theta}(\Theta) &=& 0, \qquad {\rm otherwise} \nonumber
\end{eqnarray}

The rate $\displaystyle{\frac{d {\dot N}(>\rho_0)}{dz}}$ at which we observe the inspiral
events that originate in the redshift
interval $[z, \; z + dz]$ is given by \cite{Nowak}:
\begin{eqnarray} \label{rate_nl}
\frac{d {\dot N}(>\rho_0)}{dz} &=& \frac{\dot n_0}{1+z}\; \eta(z) \; 4\pi d_M^2
\frac{d}{dz} d_M(z) \; C_{\Theta}(x) = \nonumber \\
&=& 4\pi \left( \frac{d_H}{h} \right)^3 \; \frac{\dot n_0}{1+z} \; \frac{{\bar
d}_M^2(z)}{{\cal D}(z)} \; C_{\Theta}(x)
\end{eqnarray}
where $C_{\Theta}(x) = \int_x^{\infty} P_{\Theta}(\Theta) d\Theta$ denotes the
probability that given detector registers inspiral event at redshift $z_s$ with
$\rho > \rho_0.$  The quantity $C_{\Theta}(x)$ can be calculated as
\begin{eqnarray} \label{C_theta}
C_{\Theta} (x) &=& (1+x)(4-x)^4/256 \qquad {\rm for}\;\; 0\le x \le 4 \\
               &&   0        \qquad {\rm for}\;\;    x>4   \nonumber
\end{eqnarray}
where \cite{Wang}:
\begin{eqnarray} \label{x}
x &=& \frac{4}{h\,A} (1+z)^{7/6} \left[\frac{d_A(z)}{d_H/h} \right] =
\nonumber \\
&=& \frac{4}{h\,A} (1+z)^{1/6} \; {\bar d}_M(z)
\end{eqnarray}
and
\begin{equation} \label{A}
A := 0.4733 \; \left(\frac{8}{\rho_0}\right) \left(\frac{r_0}{355\;Mpc} \right)\;
\left( \frac{{\cal M}_0}{1.2\;M_{\odot}} \right)^{5/6}
\end{equation}
Figure 1 shows the expected detection rate of inspiralling events
for the cosmological model with $(\Omega_0 = 0.3 \;, \Omega_Q =
0.7)$ assuming no source evolution and covering the full range of
a priori possible quintessential equations of states.
 It has been
obtained by numerical integration of the formula (\ref{rate_nl}).
The predictions for other realistic proportions of $\Omega_0$ and
$\Omega_Q$ are almost indistinguishable at the level of detection
rates, so the Fig.1 is representative for the whole class of
models considered. The effect of source evolution on the detection
rate is summarised in Fig.2. For transparency only one member
(corresponding to $w_q = -0.8$) of each family of curves (as in
Fig.1) is shown for different values of the evolution exponent
$D.$

The method of extracting the cosmological parameters advocated by Finn and
Chernoff \cite{Finn} makes use of the redshift distribution of observed events
in a catalogue composed of observations with the signal-to-noise ratio greater
than the threshold value $\rho_0.$ Therefore it is important to find this
distribution function for different quintessence models.
The formula for the expected distribution of observed events in the source
redshift can be easily obtained from the equation (\ref{rate_nl}):
\begin{eqnarray} \label{prob}
P(z,> \rho_0) &=& \frac{1}{{\dot N}(>\rho_0)} \frac{d {\dot N}(>\rho_0)}{dt} =
\nonumber \\
&=& \frac{ 4 \pi }{{\dot N}(>\rho_0)} \;
\left(\frac{d_H}{h}\right)^3 \; \frac{\dot n_0}{1+z}\;\eta(z) \; \frac{{\bar
d}_M^2(z)}{{\cal D}(z)} \; C_{\Theta}(x)
%
\end{eqnarray}

The summary of numerical computations for the cosmological
quintessence models considered based on the formulae (\ref{prob})
and (\ref{rate_nl}) are given in figures Fig.3 and Fig.4. Fig.3
illustrates the $P(z, >\rho_0)$ distribution function for the
$(\Omega_0 = 0.3 \;, \Omega_Q = 0.7)$ cosmological model with
different quintessential equations of state. For the purpose of
obtaining the Figures 3 and 4 we have assumed the dimensionless
Hubble constant equal to $h = 0.65$ as suggested by independent
observational evidence (e.g. SNe Ia in HST project \cite{HST} or
multiple image quasar systems \cite{QSO}). On Fig.4 the
distribution functions for different cosmological models with the
quintessence field with $w=-0.8$ have been plotted together.
Fig.5 shows the distribution functions for different evolutionary
exponent in the $(\Omega_0 = 0.3 \;, \Omega_Q = 0.7)$ model with
$w=-0.8$.

\section{Results and discussion}

It is clear from Figure 1 that different quintessential
cosmologies (singled out by $w$ parameter in the equation of
state) give different predictions for annual inspiral event rate
to be observed by future interferometric experiments.
Unfortunately , this difference is too small to be of
observational importance. Moreover, as already pointed out, there
exists a degeneracy in terms of cosmological models (labelled  by
the value of $\Omega_0$ and $\Omega_Q$). There is however a
difference between detection rates corresponding to different
values of evolutionary exponents as displayed in Figure 2.

Figure 3 shows that there is a noticeable difference in predicted
event redshift distribution functions $P(z,>\rho_0)$ for
different values of the cosmic equation of state within given
cosmological model (labelled by the values of $\Omega_0$ and
$\Omega_Q$). 
The spread between different cosmological models for
a given quintessence equation of state is much smaller as seen
from the Fig.4. This is a reflection of above mentioned effective
degeneracy with respect to values of $\Omega$ parameters.
Hopefully this degeneracy can be broken by independent estimates
of $\Omega_0$ and $\Omega_Q$ parameters in other studies (cluster
baryons estimates, Ly$\alpha$ forest surveys, large scale
structure or CMBR). The spread of redshift distribution functions
attributed to evolutionary effects is smaller as shown in Fig. 5
and has a slightly different character - the distribution
function is shifted toward increasing redshifts when the
evolutionary exponent changes from positive to negative value.
This may to some extent mimic the effect of cosmic equation of
state, but it should in principle be possible to disentangle - at
least to a certain degree from the complementary information about 
the detection rates. As can be seen from Fig.2 the magnitudes of observed event rates 
for different evolutionary exponents $D$ are clearly distinct, at least 
for the range of the Hubble constant suggested by independent 
cosmological evidence \cite{Hubble}.

The redshift distribution $P(z,>\rho_0)$ is in fact inferred from
observed chirp mass distribution. Therefore it can in principle
be distorted by the intrinsic chirp mass distribution.
Theoretical studies of the neutron star formation suggest that
masses of nascent neutron stars do not vary much with either mass
or composition of the progenitor \cite{Finn}. Also the mass
estimates of observed binary pulsars suggest that there are good
reasons to assume a negligible spread of intrinsic chirp mass (as
it was done in the present paper). Moreover any intrinsic
distribution of mass would be expected as symmetric, whereas the
redshift distribution (of cosmological origin) has certain amount
of asymmetry.

In conclusion one can hope that the catalogues of
inspiral events gathered in future gravitational waves experiments can provide
helpful information about the quintessence equation of state complementary to
that obtained by other techniques. 
Even though the most straightforward way of making inference about cosmic 
equation of state would come from future supernovae surveys 
it would be good to have in mind alternative ways of 
reaching the same goal such as the one proposed in the present paper.

\newpage

{\sc Figure Captions}

\vspace{1.5cm}

{\bf Figure 1}

The detection rate prediction for the advanced gravity wave detectors
i.e. with signal-to-noise threshold $\rho_0 = 8.$ and probing distance
$r_0=355\;Mpc.$ corresponding to quintessence cosmology with different
equations of state.

\vspace{1.5cm}
%

{\bf Figure 2}

The detection rate prediction for the advanced gravity wave
detectors i.e. with signal-to-noise threshold $\rho_0 = 8.$ and
probing distance $r_0=355\;Mpc.$ corresponding to $\Omega_0 = 0.3,
\;\; \Omega_Q = 0.7$ quintessence cosmology with $w_q=-0.8$ for
different values of evolutionary exponent $D$.

\vspace{1.5cm} {\bf Figure 3}

Redshift distribution of observed events in
the cosmological model with $\Omega_0 = 0.3, \;\; \Omega_Q = 0.7$ for
different quintessential equations of state.

\vspace{1.5cm} {\bf Figure 4}

Redshift distribution of observed events in
the cosmological quintessence model with $w=-0.8$ Different cosmological
models have been plotted collectively.

\vspace{1.5cm} {\bf Figure 5}

Redshift distribution of observed events in the cosmological
model with $\Omega_0 = 0.3, \;\; \Omega_Q = 0.7$ with $w_q =
-0.8$ quintessence for different values of evolutionary exponents
$D$.

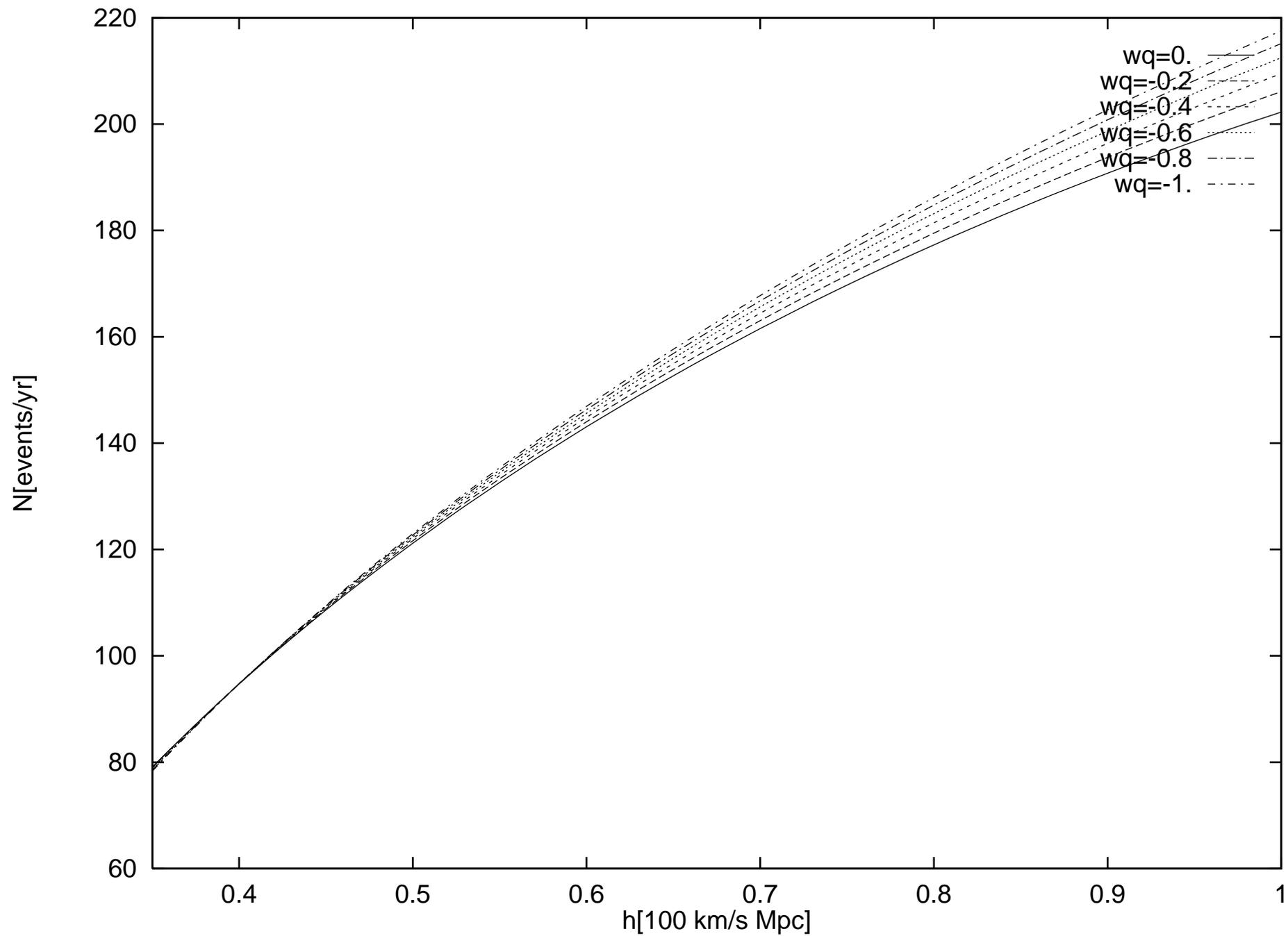
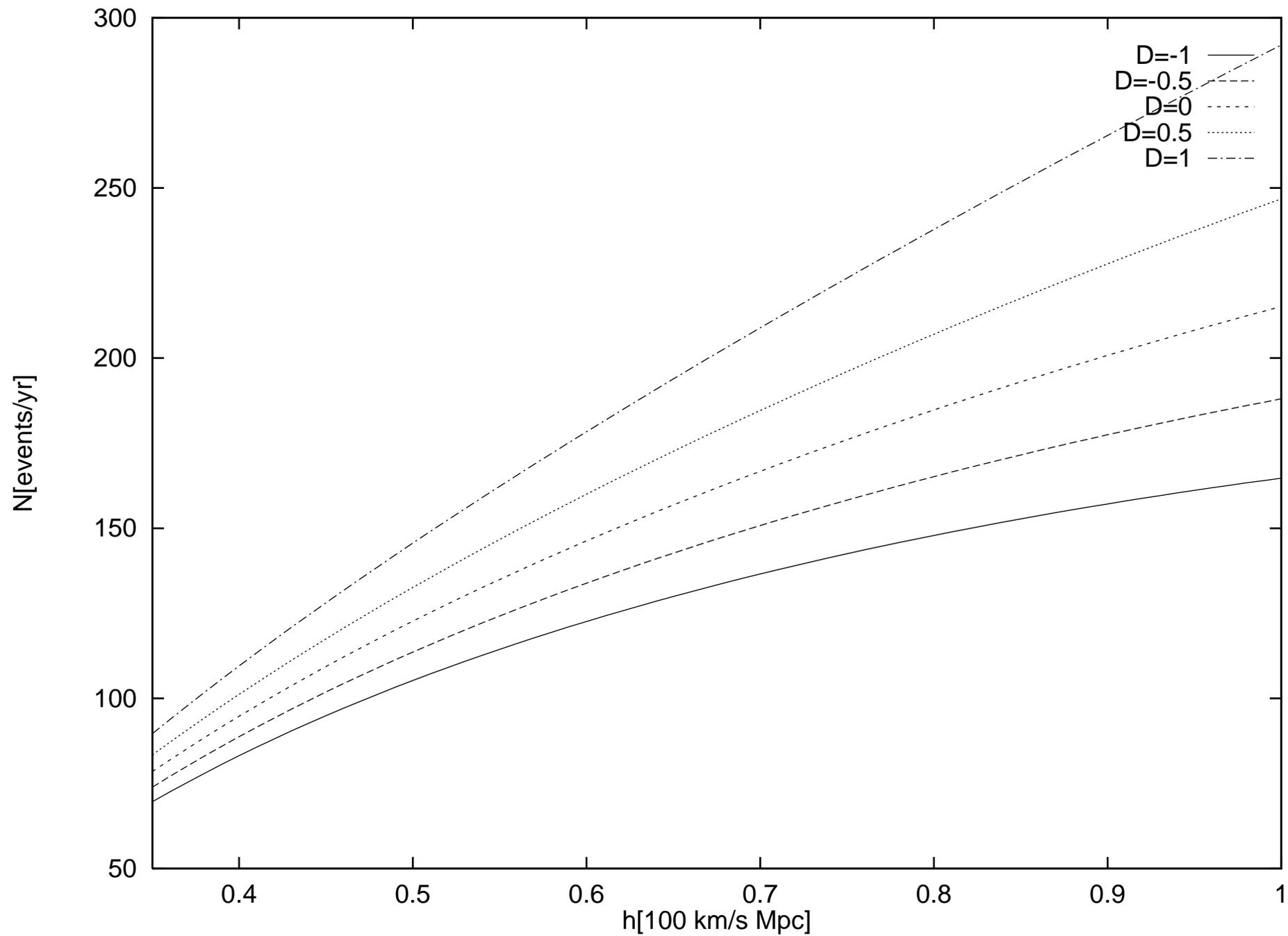
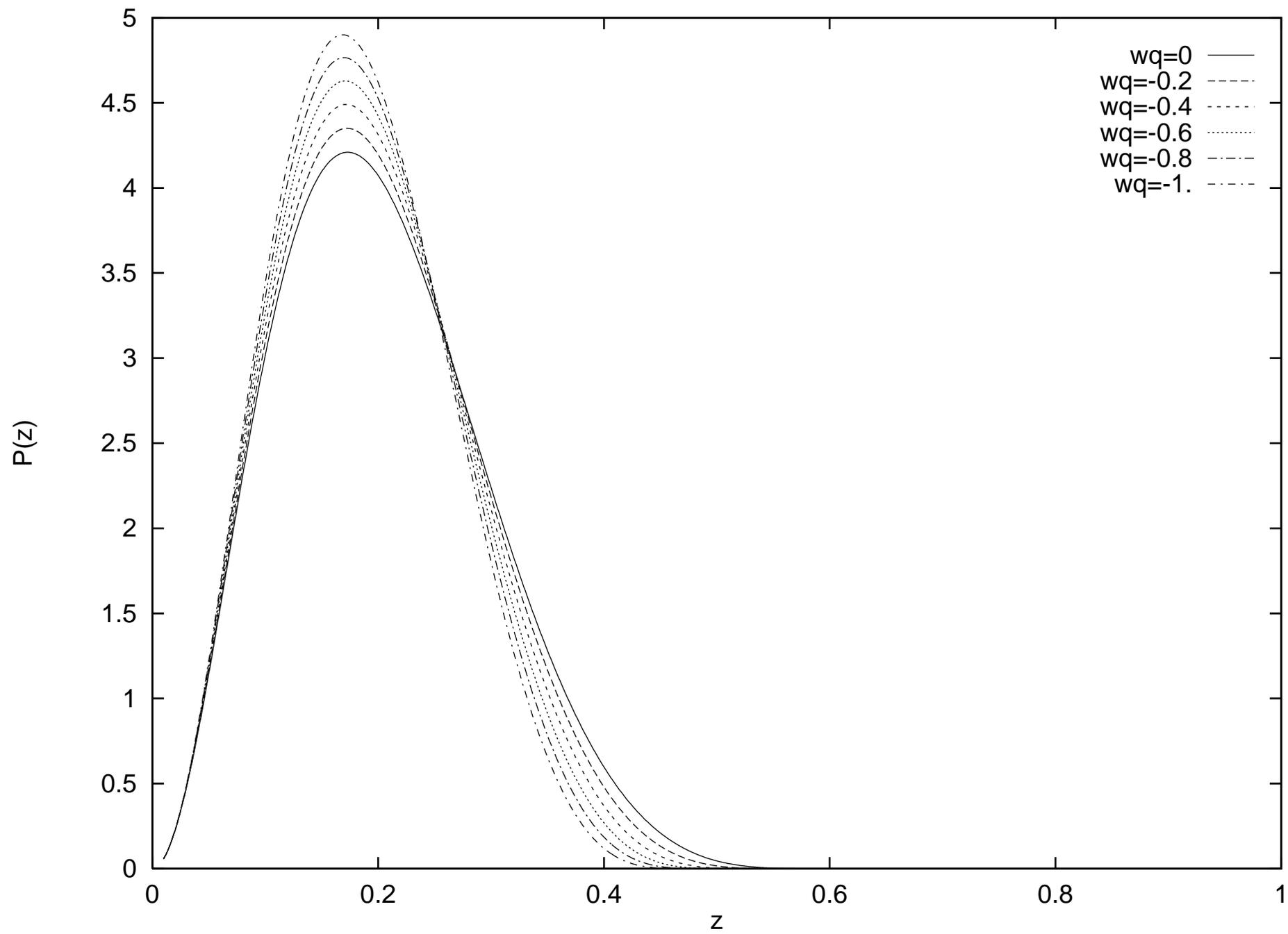
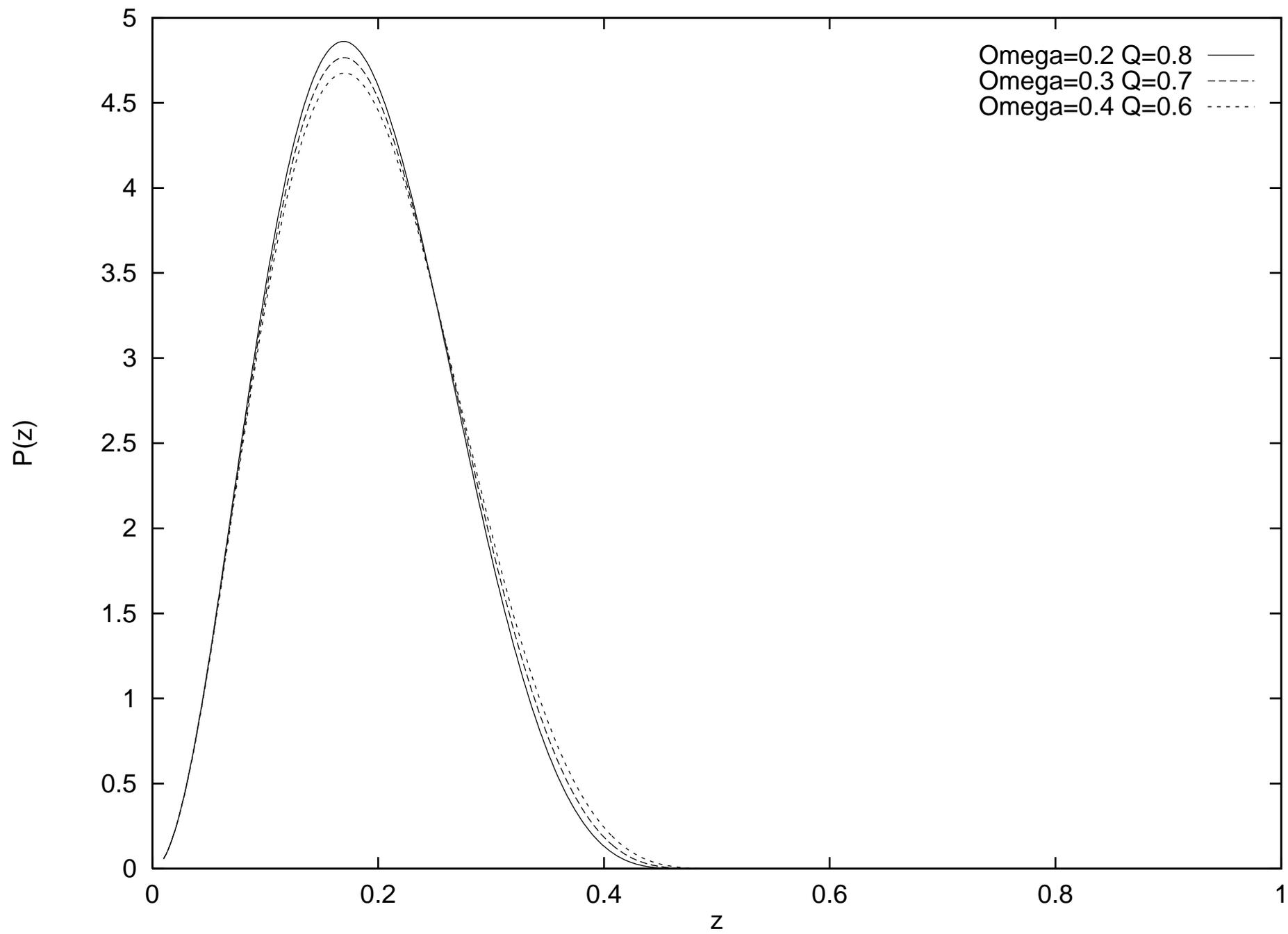
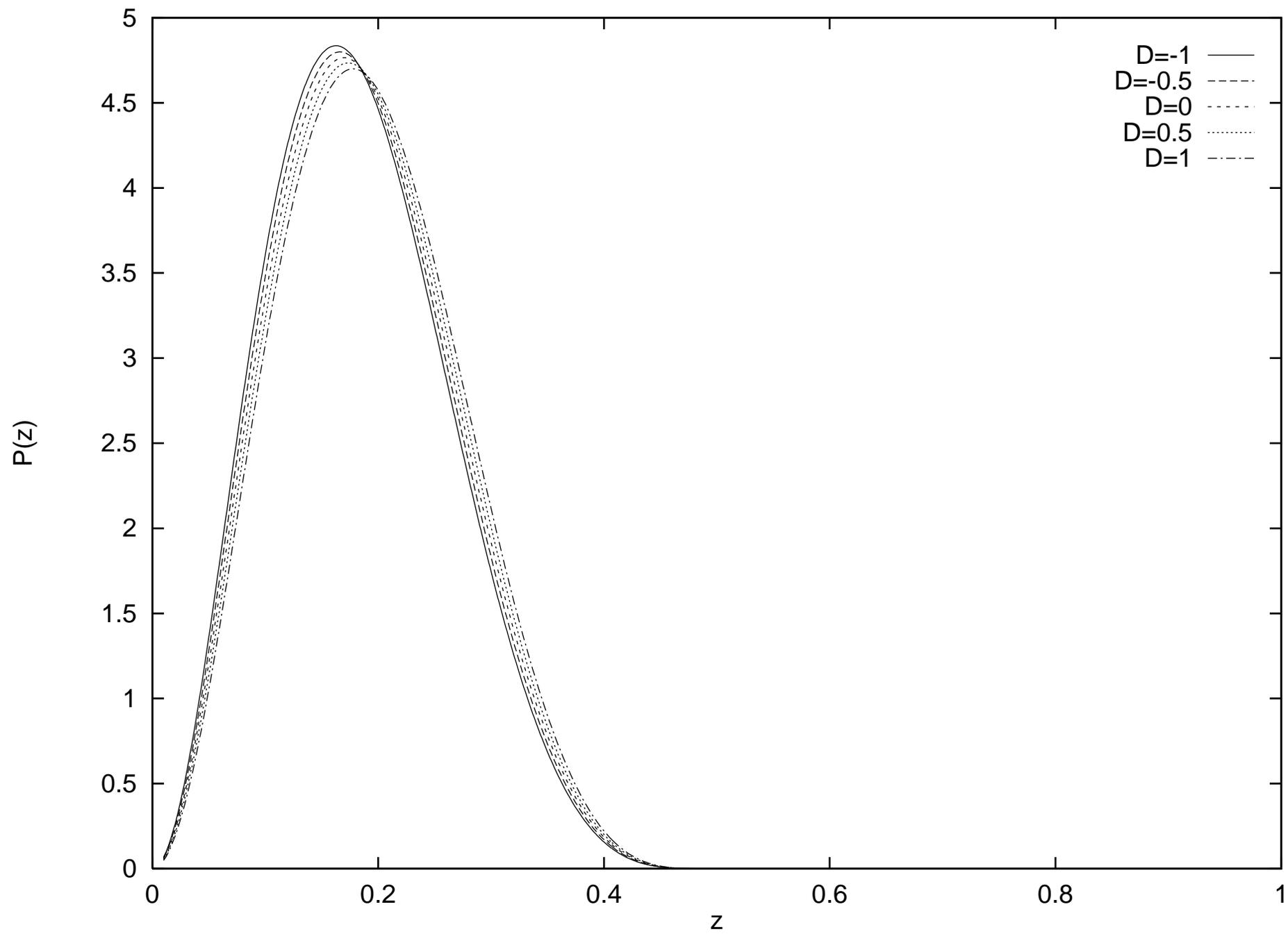
\end{document}